\documentstyle[twoside,fleqn,espcrc2,epsf]{article}
\def\fig#1#2#3{\epsfxsize=#3truein
\vskip -0.3 truein
\centerline{\epsffile{fig_#1.eps}}
\vskip 0.05 truein
\centerline{\vbox{{\bf \noindent Figure #1.} #2}}
\smallskip}

\def\figsizeA{2.4}

\def\pbp{\langle\overline{\psi}\psi\rangle}

\def\chidof{\chi^2 / {\rm d.o.f.}}
\def\MeV{{\rm\  MeV}}

\def\spose#1{\hbox to 0pt{#1\hss}}
\def\ltapprox{\mathrel{\spose{\lower 3pt\hbox{$\mathchar"218$}}
 \raise 2.0pt\hbox{$\mathchar"13C$}}}
\def\gtapprox{\mathrel{\spose{\lower 3pt\hbox{$\mathchar"218$}}
 \raise 2.0pt\hbox{$\mathchar"13E$}}}
\def\inapprox{\mathrel{\spose{\lower 3pt\hbox{$\mathchar"218$}}
 \raise 2.0pt\hbox{$\mathchar"232$}}}

\def\one{$16^4 \times 4$, $L_s=16$ and $m_0=1.9$. The
fits are to $c_0 + c_2 m_f^2$ and the stars are the $m_f=0$ extrapolated
values.}
\def\two{$\pbp$ vs. $\beta$ with
$m_0=1.9$, $m_f=0.02$ and $L_s=24$. The circles correspond
to $16^3 \times 4$ lattice with an ordered initial configuration (ic),
the crosses to a $16^3 \times 4$ with disordered ic,
the diamond to an $8^3 \times 4$ with ordered ic, and
the cross to an $8^3 \times 4$ with disordered ic.
The gauge part of the action
is a Wilson plaquette action.
}
\def\three{Same as in figure 2 but with
an Iwasaki improved gauge action with $c_1 = -0.331$.
}
\newcommand{\AmS}{{\protect\the\textfont2
  A\kern-.1667em\lower.5ex\hbox{M}\kern-.125emS}}

\title{The finite temperature QCD phase transition with domain wall fermions.}
 
\author{Pavlos M. Vranas \thanks{In collaboration with
P.~Chen,
N.~Christ,
G.~Fleming,
A.~Kaehler,
T.~Klassen,
C.~Malureanu,
R.~Mawhinney,
G.~Siegert
C.~Sui,
L.~Wu,
Y.~Zhestkov.
Supported in part by
DOE grant \# DE-FG02-92ER40699
and in part by NSF grant \# NSF-PHY96-05199 (PMV).
}
\address{Physics Dept., University of Illinois, Urbana IL 61801}
}

\begin{document}

\begin{abstract}
Results from the Columbia lattice group study of the QCD finite
temperature phase transition with dynamical domain wall fermions 
on $16^3 \times 4$ lattices are presented. These results include an
investigation of the $U(1)$ axial symmetry breaking above but close to
the transition, the use of zero temperature calculations that set the
scale at the transition and preliminary measurements close to the
transition.
\end{abstract}

\maketitle

\section{Introduction}
\label{sec:intro}

Domain wall fermions (DWF) \cite{Kaplan} provide a way to separate the
approach to the continuum limit from the approach to the chiral limit.
At fixed lattice spacing the chiral limit is approached as the size
$L_s$ of the ``fifth'' direction is increased. Furthermore, since this
regulator does not break flavor symmetry and since the computing cost
is only linear in $L_s$ it may be possible to simulate two flavor QCD
with three light pions close to the finite temperature transition
using today's supercomputers.

Preliminary results of QCD thermodynamic studies with DWF were
presented in \cite{lat98_vranas}. Results from simulations on $8^3
\times 4$ lattices indicated the presence of a low temperature phase
with spontaneous chiral symmetry breaking but intact flavor symmetry
and a high temperature phase with the full $SU(2) \times SU(2)$ chiral
flavor symmetry.  Also, preliminary results on $16^3 \times 4$
lattices just above the deconfining transition indicated a difference
in the $\pi$ and $\delta$ susceptibilities and screening masses. This was
preliminary evidence for the anomalous breaking of the $U_A(1)$
symmetry above the transition.

Here the above preliminary results are extended in two directions: 1)
The investigation of the $U_A(1)$ axial symmetry breaking is completed
with results at two values of the coupling close to the transition.
2) A full study of the transition region on $16^3 \times 4$ lattices
is performed and is accompanied by scale setting calculations at the
transition point to determine the transition temperature and the pion
mass. Some of these results have already been reported in
\cite{dpf99_vranas}. Also see \cite{lat99_rdm,lat99_karsch}.

The DWF action used in this work is as in \cite{Furman_Shamir} with
the modifications as in \cite{PMV}. Some details on the numerical
methods can be found in \cite{lat98_vranas}. For a review on DWF
see \cite{lat98_blum} and references therein.

\section{The $U_A(1)$ symmetry above $T_c$}
\label{sec:u1}

An important question concerns the role of the $U_A(1)$ anomaly above the
finite temperature QCD transition. 
Earlier attempts to address this question \cite{omega,Kogut_Lagae_Sinc}
using staggered fermions could not produce conclusive results because
staggered fermions break the $U_A(1)$ symmetry already at the
classical level.
On the other hand, DWF bring this problem under control. In particular
at $L_s= \infty$ massless DWF do not break the $U_A(1)$ symmetry at the
classical level. Also, 
the DWF Dirac operator can have exact zero modes \cite{NN1}, the
necessary property for anomalous breaking. In \cite{CU_zero_modes} it
was shown that this property is maintained to a good approximation
for fairly small masses and for $L_s$ as small as $\approx 10$.

The difference of the susceptibilities $\chi$ of the $\pi$ and $\delta$ is
used as a measure of anomalous symmetry breaking.  This difference was
measured as a function of the bare quark mass $m_f$ on a $16\times 4$
lattice at $\beta=5.45$ and $\beta=5.40$ ($\beta_c \approx 5.325$).
The results are presented in figure 1.  The lines are fits to $c_0 +
c_2 m_f^2$ and have $\chidof \approx 1$. The absence of a linear term
indicates that for $L_s=16$ the chiral symmetry is effectively
restored. The $m_f=0$ extrapolated values are $0.26(6)$
at $\beta=5.40$ and $0.08(3)$ at $\beta=5.45$.  Although both are not
zero by a statistically significant amount
their value is small when compared with $\chi_\pi$ and $\chi_\delta$
which are $\approx 8$.  Universality arguments require that if the
QCD phase transition is to be second order, the anomalous $U(1)_A$
must be broken.  It is an open question as to whether the small size
of the $U(1)_A$ symmetry breaking seen here is sufficient to support
this theoretical prediction that the two-flavor QCD phase transition
is second order \cite{Pizarski}.  For quenched QCD thermodynamic 
studies with DWF see 
\cite{lat98_fleming,lat98_kaehler,lat99_Sinc},
and with overlap fermions see \cite{lat99_Heller}.

\fig{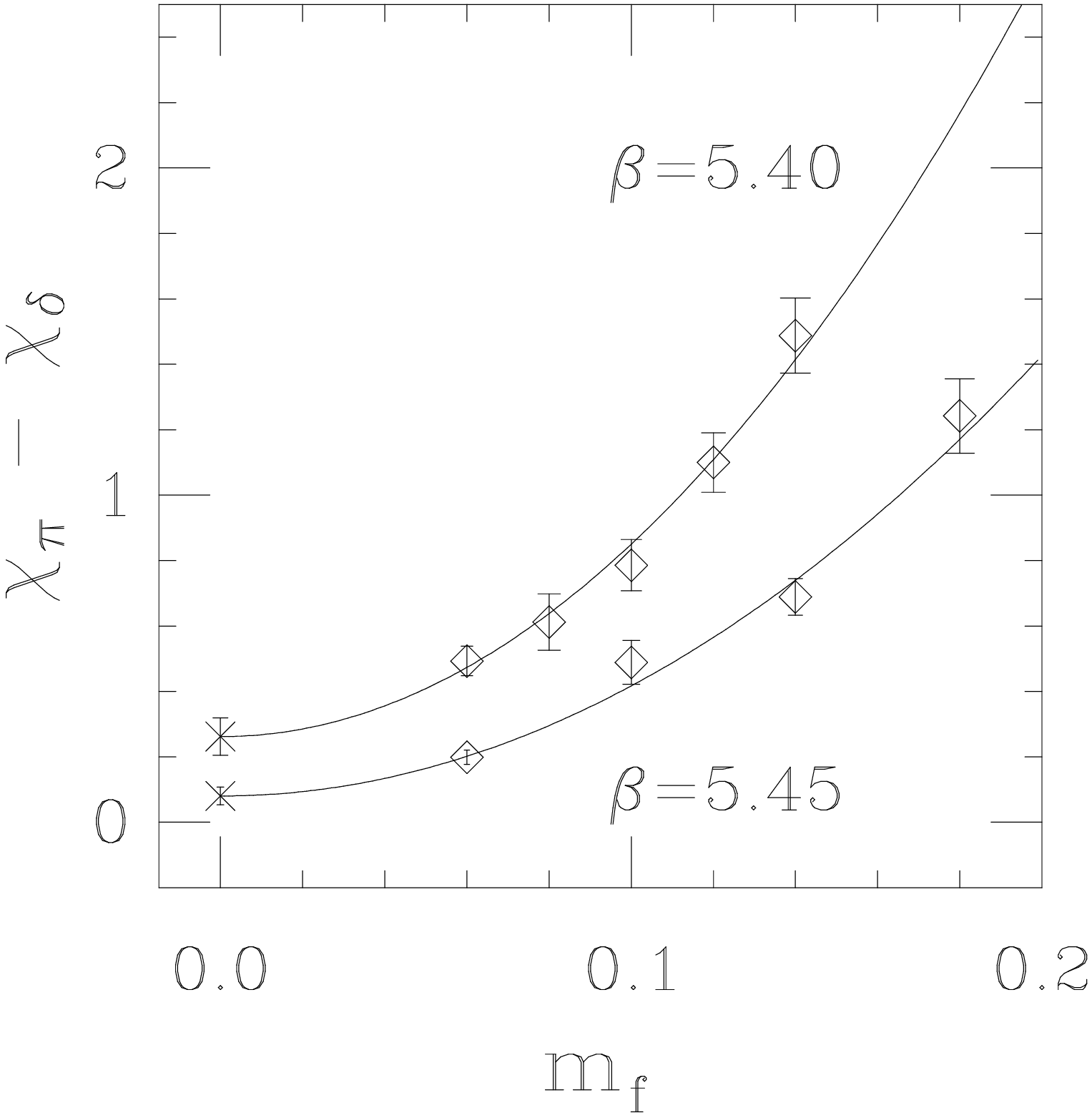}{\one}{\figsizeA}

\section{The transition}
\label{sec:transition}

The transition region ($5.2 < \beta < 5.4$) is studied on $16^3
\times 4$ lattices using $m_f=0.02$, a domain wall
height $m_0 = 1.9$ and $L_s=24$. These values were chosen based on
previous exploratory studies on smaller lattices 
\cite{lat98_vranas,dpf99_vranas}. 
In particular $m_0=1.9$ was found to produce two
flavor physics and $L_s=24$ to be large enough to keep the
residual chiral symmetry breaking effects on the chiral condensate to
the few percent level above the transition and to $\approx 15\%$ level
below the transition.  The results are shown in figure 2. The
gauge part of the action is a standard Wilson plaquette action.

In order to set the scale a simulation \cite{lat99_wu}
on an $8^3 \times 32$ lattice
was done at $\beta=5.325$ corresponding to the middle of the transition
region . It was found that in lattice units $m_\rho = 1.18(3)$ 
and $m_\pi = 0.654(3)$. This gives a value for the
critical temperature $T_c = 163(4) \MeV$ and $m_\pi = 427(11) \MeV$.
The critical temperature is in agreement with results obtained from
other fermion regulators \cite{lat99_karsch}. The pion mass is clearly
too heavy to be able to extract useful information regarding the
order of the transition. More sophisticated studies
\cite{lat99_fleming} indicate that the residual chiral symmetry
breaking effects are much larger than expected and in order to obtain
a physical pion mass $L_s \approx 100$ may be needed.

\fig{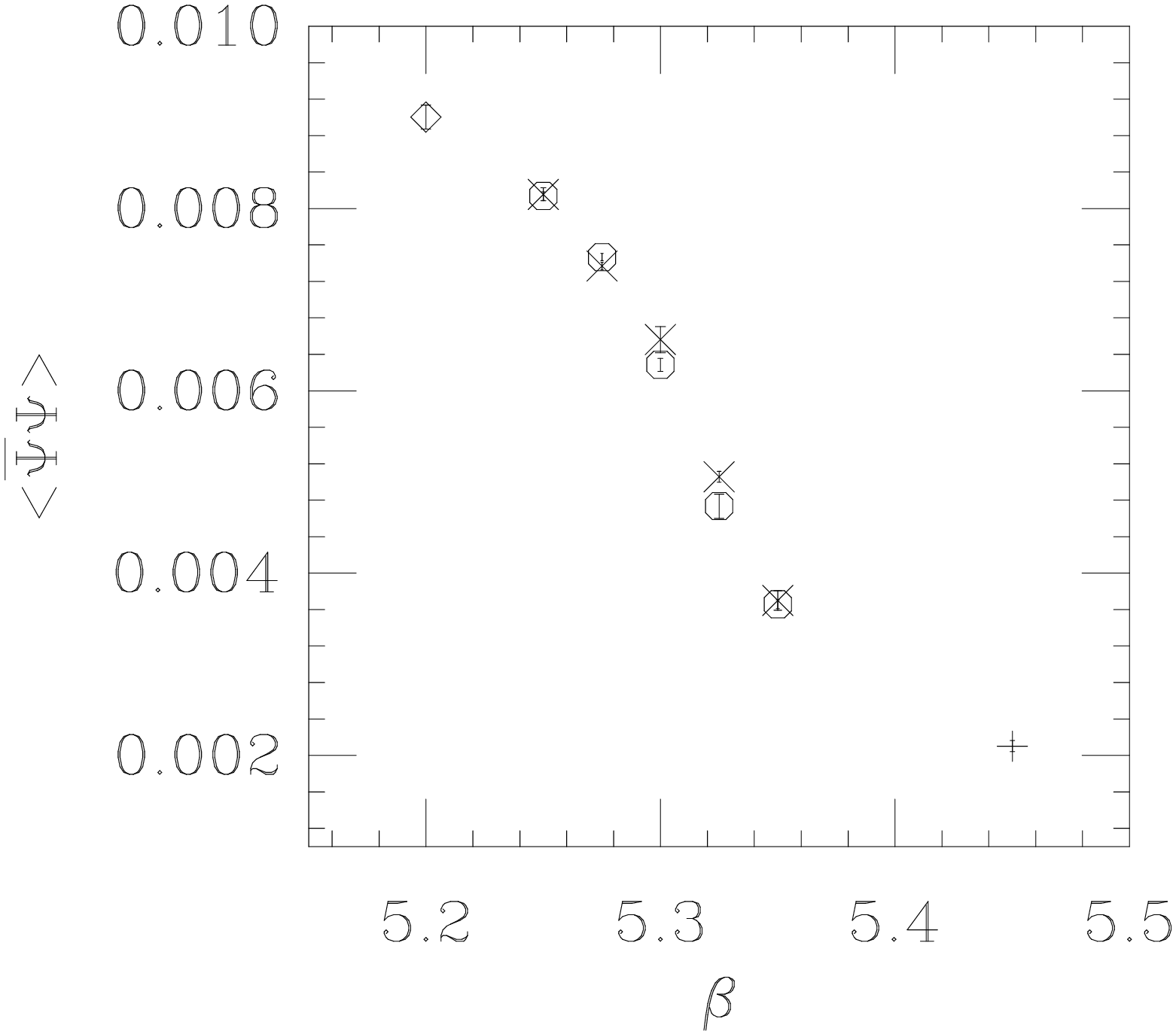}{\two}{\figsizeA}

Large residual chiral symmetry breaking effects have been related to
the ``roughness'' of the gauge field configurations
\cite{NN1,Furman_Shamir,PMV,EHN_flow}.  Therefore a possible ``cure''
could be to use improved gauge actions. This idea was explored in zero
temperature quenched QCD by using an Iwasaki improved gauge
action.  Dramatic improvements were seen at couplings that correspond
to $N_t=4$ finite temperature quenched QCD just above the transition
\cite{lat99_wu}. These results prompted a study of the dynamical QCD
finite temperature transition using Iwasaki improved gauge action
with $c_1 = -0.331$. The results are presented in figure 3. All
parameters are the same as in figure 2.

\fig{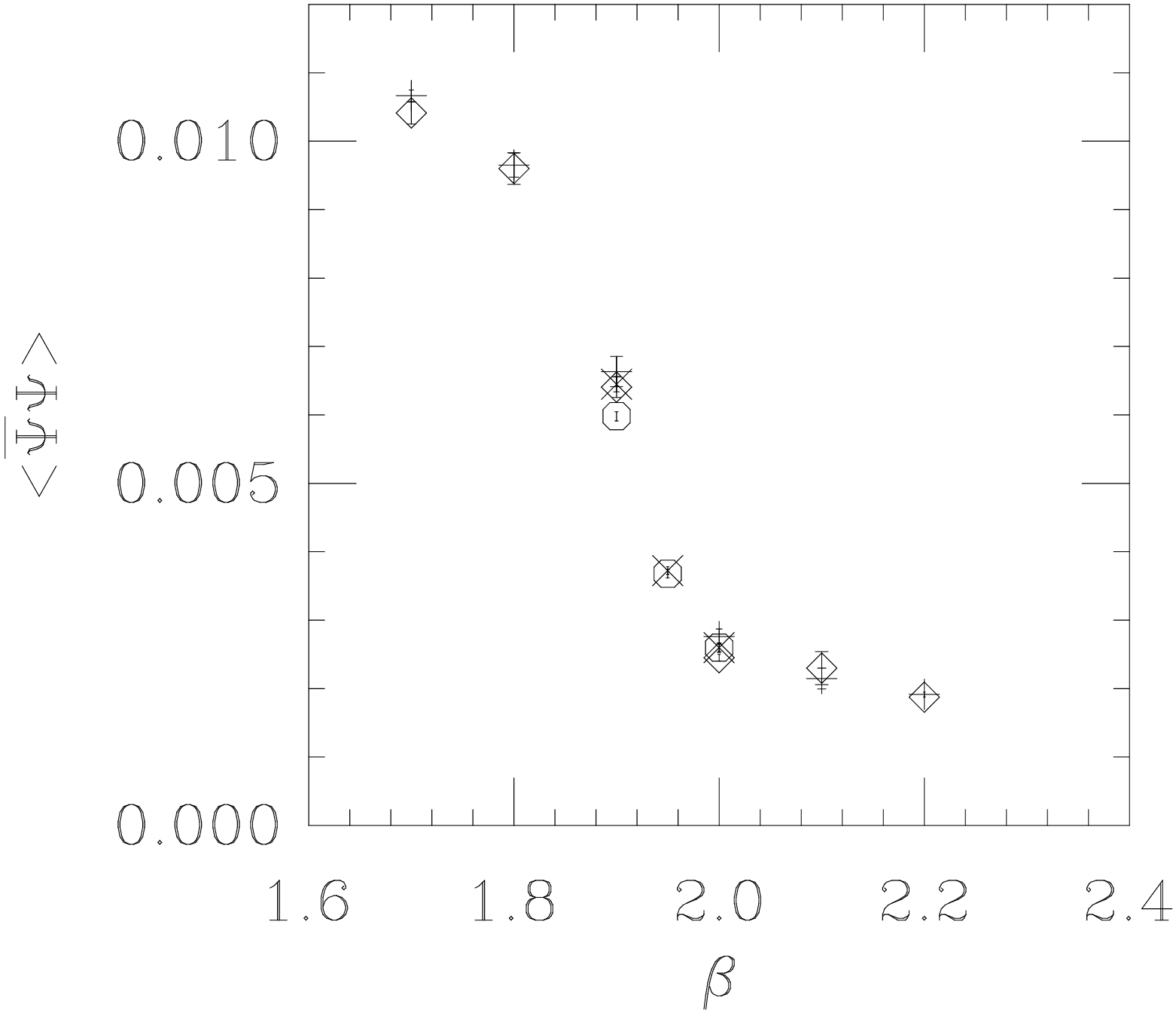}{\three}{\figsizeA}

Again, in order to set the scale a simulation \cite{lat99_wu}
on an $8^3 \times 32$
lattice was done at $\beta=1.9$ corresponding to the middle of the
transition region. It was found that in lattice units
$m_\rho = 1.163(21)$ and $m_\pi = 0.604(3)$. This gives a value for
the critical temperature $T_c = 166(3) \MeV$ and $m_\pi = 400(7) \MeV$.  
The critical temperature is in agreement with the
Wilson plaquette action results. However, the pion mass
did not get lighter by a significant amount.

\section{Conclusions}
\label{sec:conclusions}

The difference of the susceptibilities of the $\pi$ and $\delta$ was
measured at two couplings just above the finite temperature
transition.  This difference was found to be non-zero by a
statistically significant amount indicating that $U_A(1)$ remains
broken above but close to the transition.  However, the value of the
difference  of the susceptibilities 
is much smaller than the susceptibilities themselves (by a
factor of $\approx 30 - 100$ for each point) suggesting that the size
of $U_A(1)$ is indeed rather small at these couplings. It is an open
question as to what effect this may have to the order of the
transition.

The transition region was studied by dynamical simulations on $16^3
\times 4$ lattices with $L_s=24$ and the critical coupling was
located. At that coupling zero temperature simulations were done on
$8^3 \times 32$ lattices in order to set the scale. For the gauge part
of the action two different types were used.  A standard plaquette
action gave critical temperature 
$T_c = 163(4) \MeV$ and $m_\pi = 427(11) \MeV$.
In an effort to reduce the pion mass an Iwasaki gauge
action was also used. For that action $T_c = 166(3) \MeV$ and $m_\pi =
400(7) \MeV$.  Unfortunately 
the pion mass is still too heavy to allow a study of
the order of the transition. Larger values of $L_s$ or alternative
improvement techniques must be employed in order to reduce the pion
mass to physical values and still be able to study the transition with
the available computing resources.

All calculations were done on the $400$ Gflops QCDSP machine at
Columbia University.


\begin{thebibliography}{9}

\bibitem{Kaplan} D.B. Kaplan, Phys. Lett. {\bf B288} (1992) 342.

\bibitem{lat98_vranas} P. Chen et.al., Nucl. Phys. {\bf B73}
(Proc. Suppl.) (1999) 456.

\bibitem{dpf99_vranas} P. Vranas, DPF 99 proc., hep-lat/9903024.

\bibitem{lat99_rdm} R. Mawhinney, these proceedings.

\bibitem{lat99_karsch} F. Karsch, these proc., hep-lat/9909006.

\bibitem{Furman_Shamir} V. Furman, Y. Shamir, Nucl. Phys. {\bf B439} (1995) 54.

\bibitem{PMV} P.M. Vranas, Phys. Rev. {\bf D57} (1998) 1415.

\bibitem{lat98_blum} T. Blum, Nucl. Phys. {\bf B73}
(Proc. Suppl.) (1999) 167.

\bibitem{omega} S. Chandrasekharan et.al., 
Phys. Rev. Lett. {\bf 82} (1999) 2463.

\bibitem{Kogut_Lagae_Sinc}
J.B. Kogut et.al.,
Phys. Rev. {\bf D58} (1998) 054504.

\bibitem{NN1} 
R. Narayanan, H. Neuberger, Nucl. Phys. {\bf B443} (1995) 305.

\bibitem{CU_zero_modes} P. Chen et.al., Phys. Rev. {\bf D59} (1999) 054508.

\bibitem{Pizarski} R. Pizarski and F. Wilczek, Phys. Rev. {\bf D29} 
(1984) 338.

\bibitem{lat98_fleming} P. Chen et.al., Nucl. Phys. {\bf B73}
(Proc. Suppl.) (1999) 207.

\bibitem{lat98_kaehler} P. Chen et.al., Nucl. Phys. {\bf B73}
(Proc. Suppl.) (1999) 405.

\bibitem{lat99_Heller} U. Heller, these proc., hep-lat/9908036.

\bibitem{lat99_Sinc} D. Sinclair, these proceedings.


\bibitem{lat99_fleming} G. Fleming, these proceedings.

\bibitem{EHN_flow}
U. Heller et.al.,
Nucl. Phys. {\bf B535} (1998) 403.

\bibitem{lat99_wu} L. Wu, these proceedings.


\end{thebibliography}
\end{document}